\documentclass[aps,prb,reprint,superscriptaddress]{revtex4-2}
\usepackage{graphicx}
\usepackage{soul}
\usepackage{xcolor}
\usepackage{amsmath}

\begin{document}


\title{Enhanced magneto-optical intensity effect in a helicity-preserving all-dielectric metasurface at Mie resonances and the anapole state}

\author{P.V. Zorina}
\email[]{p.zorina@rqc.ru}
\affiliation{Russian Quantum Center, 121205 Moscow, Russia}
\affiliation{Moscow Institute of Physics and Technology (MIPT), 141700 Dolgoprudny, Russia}

\author{D.O. Ignatyeva}
\affiliation{Russian Quantum Center, 121205 Moscow, Russia}
\affiliation{Faculty of Physics, M.V. Lomonosov Moscow State University, Moscow, Russia}

\author{A.E. Bezmenova}
\affiliation{Russian Quantum Center, 121205 Moscow, Russia}

\author{A.N. Kalish}
\affiliation{Russian Quantum Center, 121205 Moscow, Russia}
\affiliation{Faculty of Physics, M.V. Lomonosov Moscow State University, Moscow, Russia}

\author{S. Xia}
\affiliation{National Engineering Research Center of Electromagnetic Radiation Control Materials, University of
Electronic Science and Technology of China, Chengdu 610054, China}
\affiliation{State Key Laboratory of Electronic Thin-Films and Integrated Devices, University of Electronic Science and
Technology of China, Chengdu 610054, China}

\author{L. Bi}
\affiliation{National Engineering Research Center of Electromagnetic Radiation Control Materials, University of
Electronic Science and Technology of China, Chengdu 610054, China}
\affiliation{State Key Laboratory of Electronic Thin-Films and Integrated Devices, University of Electronic Science and
Technology of China, Chengdu 610054, China}

\author{V.I. Belotelov}
\affiliation{Russian Quantum Center, 121205 Moscow, Russia}
\affiliation{Faculty of Physics, M.V. Lomonosov Moscow State University, Moscow, Russia}

\date{\today}

\begin{abstract}
Nanophotonic structures provide an efficient route to enhancing magneto-optical effects by concentrating electromagnetic fields at subwavelength scales. In this work, we propose and experimentally demonstrate a helicity-preserving all-dielectric metasurface for enhancing magnetization-induced transmission modulation under circularly polarized excitation. The optical response of the structure is governed by the Mie resonances of silicon nanodisks and by a spectral feature associated with the anapole state. In the spectral regions of the Mie resonances, the metasurface exhibits a pronounced enhancement of the magneto-optical intensity response relative to a bare magnetic film of the same thickness, with the normalized magneto-optical intensity effect increased by a factor of about 2--3. A strong response is also observed in the spectral region associated with the anapole state, manifested as a local transmission maximum. In this regime, the normalized magneto-optical intensity effect exceeds that of the bare magnetic film by about 30\%, while the metasurface transmission remains as high as around 80\%. The enhanced response in this spectral region is preserved over a broad range of incidence angles. These results demonstrate that all-dielectric metasurfaces combining Mie resonances with the spectral feature associated with the anapole state provide an efficient platform for magneto-optical intensity modulation of circularly polarized light at high transmission.
\end{abstract}


\maketitle

\section{Introduction}
Magneto-optical effects are widely used for light modulation~\cite{kharratian2019rgb, kharratian2020advanced}, including non-reciprocal classic optic and quantum devices \cite{ren2022single, blumenthal2024enabling, cao2017efficient, portela2023novel, pandey2026highly}. The speed of such a modulation can be very high and reach up to THz frequencies~\cite{grebenchukov2021terahertz, shah2023magneto, blank2023magneto}. However, the efficiency of photon-spin interaction in bulk magneto-optical materials is often limited, typically requiring large crystal volumes or high magnetic fields to achieve substantial effects. A powerful alternative to overcome this limitation lies in the utilization of nanostructured and resonant photonic designs. 

By engineering optical resonances in subwavelength structures such as dielectric nanoparticles, metasurfaces, or photonic crystals, the local electromagnetic field can be drastically enhanced and confined. This approach leverages a broad set of localized and collective modes, including plasmonic resonances in metals, magnetic and electric Mie resonances in high-index dielectrics, high-quality-factor Fabry--Perot cavity modes, guided-mode resonances, and bound states in the continuum (BIC) in periodic metasurfaces~\cite{maccaferri2020nanoscale}. Representative demonstrations of resonantly enhanced magneto-optical responses include hybrid dielectric--ferromagnetic nanoresonators based on magnetic Mie modes~\cite{barsukova2017magneto, mamian2025tailoring}, tunable all-dielectric metasurfaces operating at electric and magnetic dipole resonances~\cite{zorina2025thermally}, magneto-optical gratings exploiting high-$Q$ quasi-BIC states~\cite{tang2025giant, nazarenko2025bic}, magnetically tunable intrinsically chiral photonic crystal slabs supporting chiral quasi-BICs with near-unity circular dichroism and ultra-high Q-factors~\cite{yang2025magnetically}, one-dimensional magnetic garnet gratings supporting guided-mode resonances with two-orders-of-magnitude enhancement of the magneto-optical response~\cite{voronov2020magneto}, and hybrid magneto-plasmonic crystals combining surface plasmon polaritons and guided modes~\cite{kolmychek2015magneto, murzina2025waveguide}. Additional functionality can be achieved by coupling distinct resonant states, for example by combining a Mie resonance with a high-$Q$ waveguide or cavity mode~\cite{yang2025giant}, or by hybridizing multipolar dark plasmonic modes in magnetoplasmonic nanocavities~\cite{lopez2020enhanced}. Beyond pure enhancement, complex modal engineering may also enable qualitatively new effects, such as anomalous polarization modulation and rotation associated with magnetic-mode excitation in silicon--garnet resonators~\cite{xia2022circular}, as well as broadband polarization-multiplexed nonreciprocity at the device level~\cite{cao2025broadband}.

\begin{figure*}[htb]
\centering
(a)~~~~~~~~~~~~~~~~~~~~~~~~~~~~~~~~~~~~~~~~~~~~~~~~~~
(b)~~~~~~~~~~~~~~~~~~~~~~~~~~~~~~~~~~~~~~~~~~~~~~~~~
(c) \\ 
\includegraphics[width=0.3\linewidth]{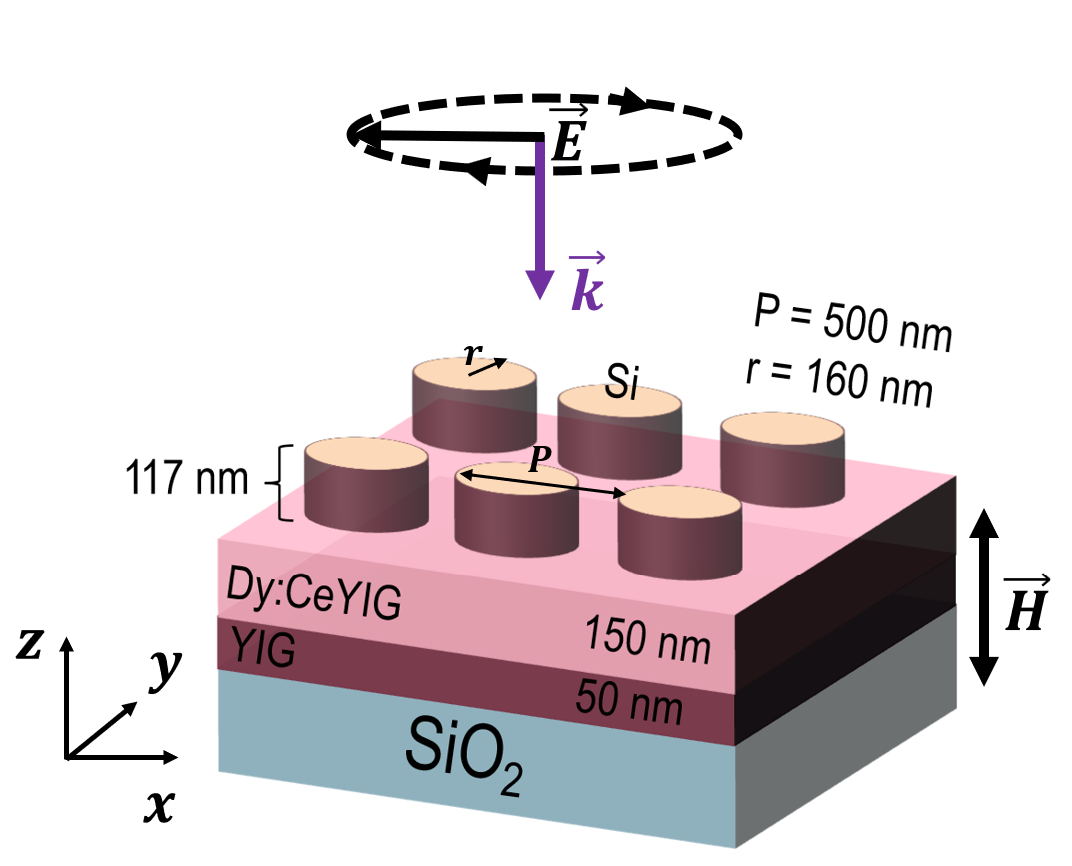}
\includegraphics[width=0.33\linewidth]{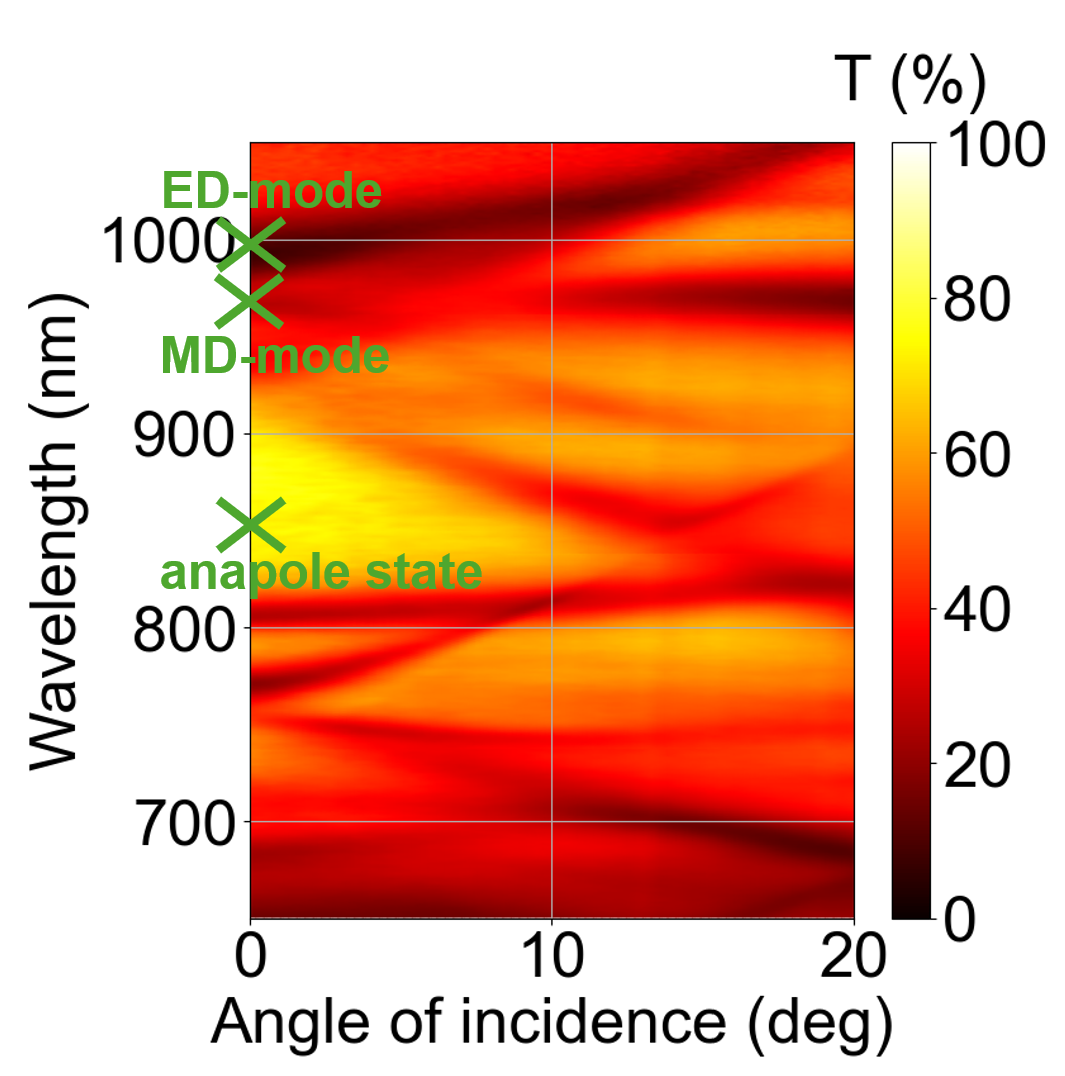}
\includegraphics[width=0.31\linewidth]{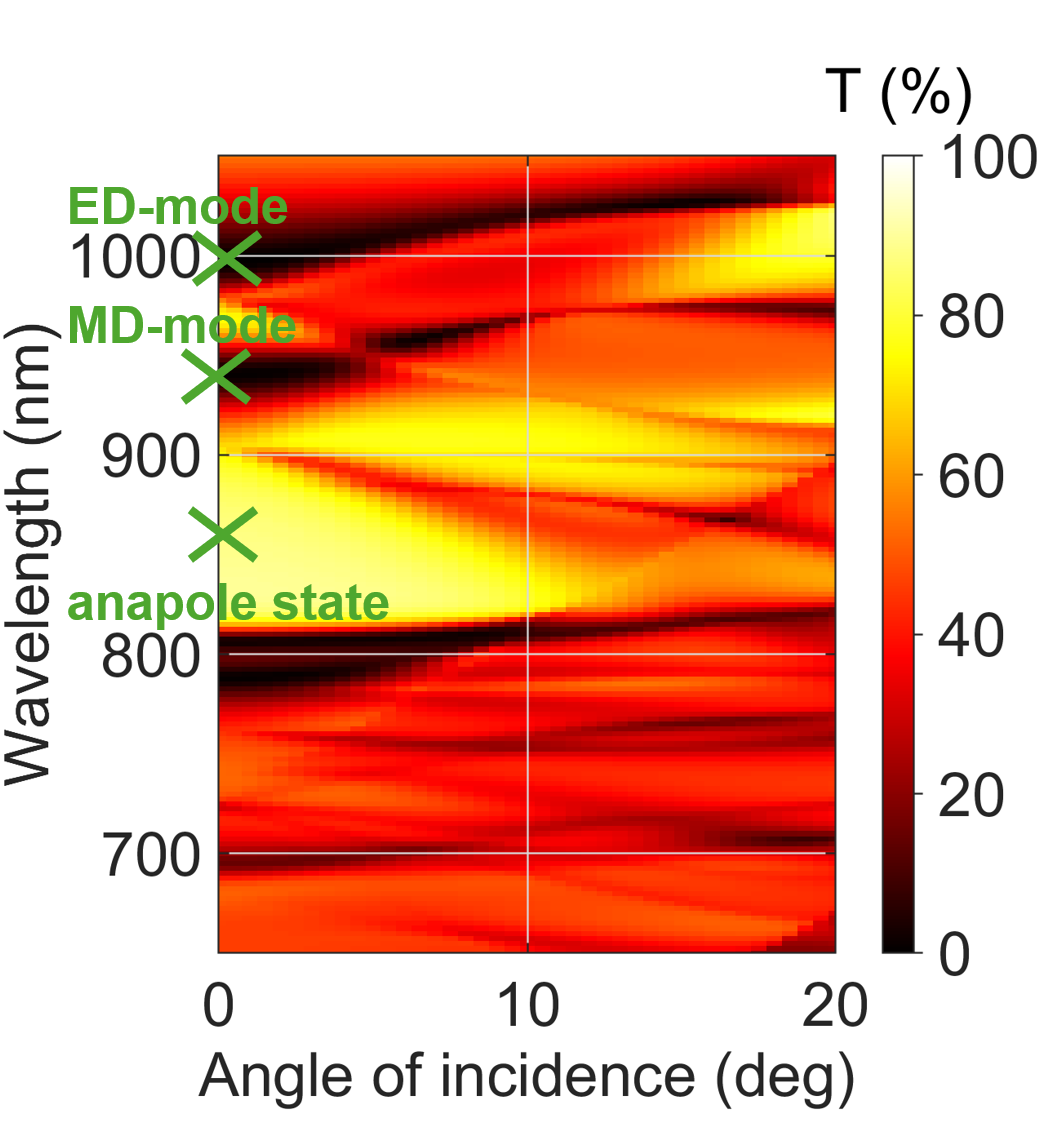}
\caption{(a) Scheme of the investigated metasurface. (b,c) Angle-resolved (b) experimental and (c) simulated transmission spectra of the sample under circularly polarized excitation. Green marks denote the positions of the discussed ED, MD modes and anapole state.}
\label{Fig: scheme and T}
\end{figure*}

One of the most rapidly expanding and promising areas of current research is all-dielectric nanophotonics, which focuses on the study of high-refractive-index semiconductor and dielectric nanoparticles (such as Si, Ge)~\cite{gurvitz2019high, kruk2017functional}. Their ability to support electric and magnetic multipole resonances enables efficient control of light at the nanoscale, and multipole expansion provides the standard framework for describing light scattering by such particles~\cite{masoudian2022multifaceted}.

An important concept in all-dielectric nanophotonics is the anapole state, commonly regarded as one of the simplest realizations of a nonradiating electromagnetic source, originally introduced by Zeldovich~\cite{zel1959parity}. In nanophotonics, the anapole state can be understood as destructive far-field interference between the electric dipole (ED) and the toroidal electric dipole (TED), which possess the same radiation symmetry; their superposition can therefore strongly suppress scattering~\cite{savinov2014toroidal, wei2016excitation, li2018origin, monticone2019can}. Experimentally, dynamic anapole excitation was first demonstrated in the microwave regime~\cite{fedotov2013resonant} and subsequently observed at optical frequencies in a single silicon nanodisk~\cite{miroshnichenko2015nonradiating}. A hallmark of the anapole regime is the simultaneous suppression of far-field radiation and strong electromagnetic energy localization inside the nanostructure~\cite{yang2018anapole, zenin2017direct}. From a magneto-optical perspective, such field localization is expected to increase the effective light--matter interaction within the magnetic material, while reduced radiative losses make the anapole state a promising regime for enhancing magneto-optical effects.

Among such effects, magneto-optical transmission modulation is of particular interest, since it directly affects the transmitted intensity of light in the presence of magnetization. In the Faraday geometry, the magneto-optical activity of the material leads to a change in the transmission coefficient for circularly polarized light when the magnetization direction is reversed. Such intensity-type magneto-optical responses are closely related to polarization-dependent absorption effects and have long been used to probe magnetically induced changes in optical transitions~\cite{kimel20222022,mack2007application}. Recent advances in nanophotonics have shown that magneto-optical intensity effects can be strongly enhanced by photonic resonances in plasmonic~\cite{uchiyama2025plasmon} and dielectric~\cite{mase2025dielectric} nanostructures, shifting the paradigm from composition-defined to structure-controlled responses.

In this work, we investigate the enhancement of magneto-optical intensity effectn in an all-dielectric magneto-optical metasurface at Mie resonances and at the anapole state. Particular attention is paid to these two spectral features because of their strong field confinement and their high potential for boosting light--matter interaction in magnetic nanophotonic structures. Our aim is to relate the spectral-angular distribution of the transmission-modulation signal to the optical modes of the structure and to compare experimental results with numerical modelling and theoretical analysis.


\section{Results}
\subsection{Optical modes of the metasurface}
We consider the structure schematically shown in Fig.~\ref{Fig: scheme and T}a. The magnetic garnet films were fabricated on a fused-silica ($\mathrm{SiO_2}$) substrate. A two-dimensional square array of silicon nanodisks with period $P=500$~nm was then fabricated, with nanodisk radius $r=160$~nm and height $h=117$~nm. The fabrication procedure is described in Appendix~A, the experimental setup is presented in Appendix~B, and the optical and magneto-optical material parameters are given in Appendix~C. Due to the high refractive index of silicon, the nanodisks support pronounced Mie-type resonances, including electric and magnetic dipole modes, together with the anapole state. These modes provide the spectral conditions under which the magneto-optical response of the garnet layer is enhanced.

\begin{figure*}[t]
\centering

(a)\includegraphics[width=0.305\linewidth]{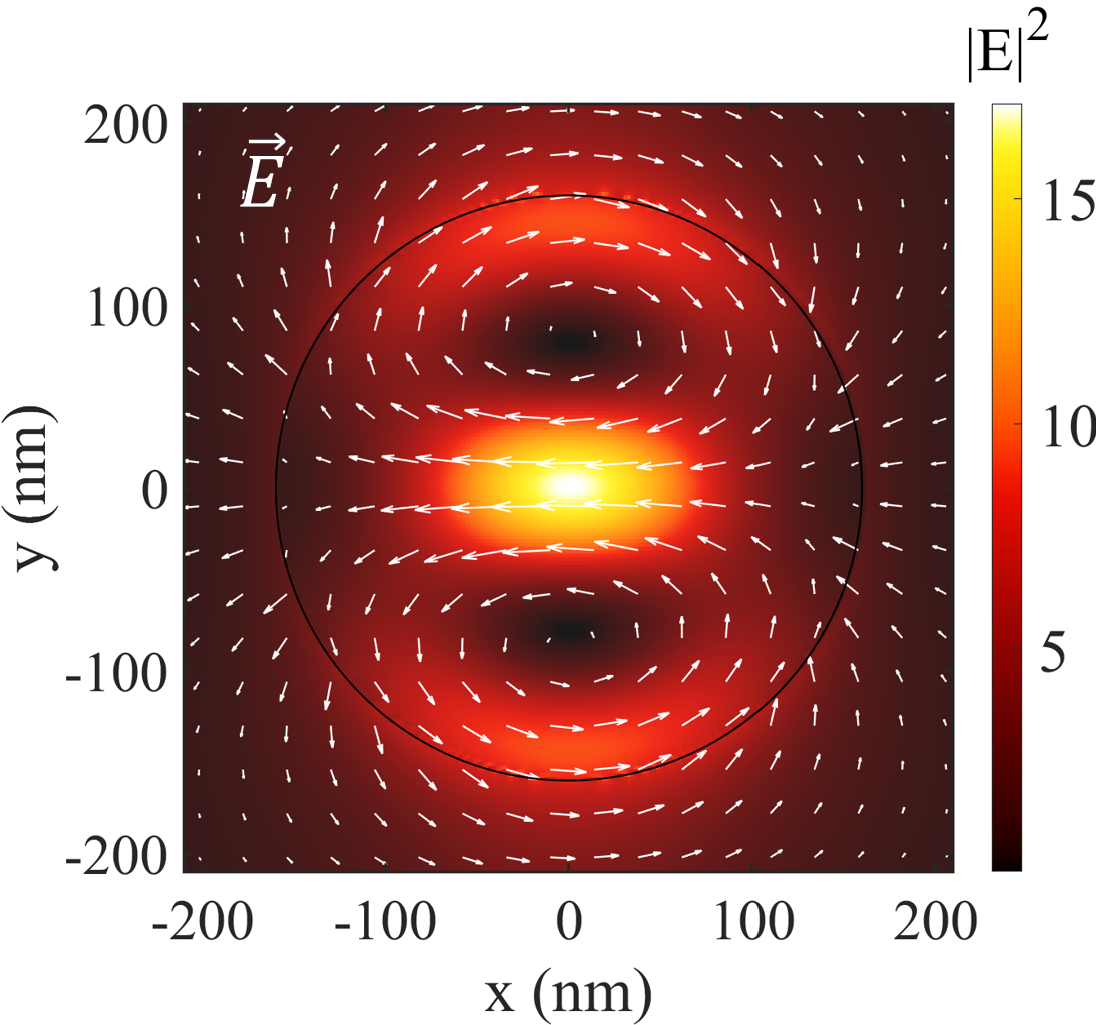}
(b)\includegraphics[width=0.306\linewidth]{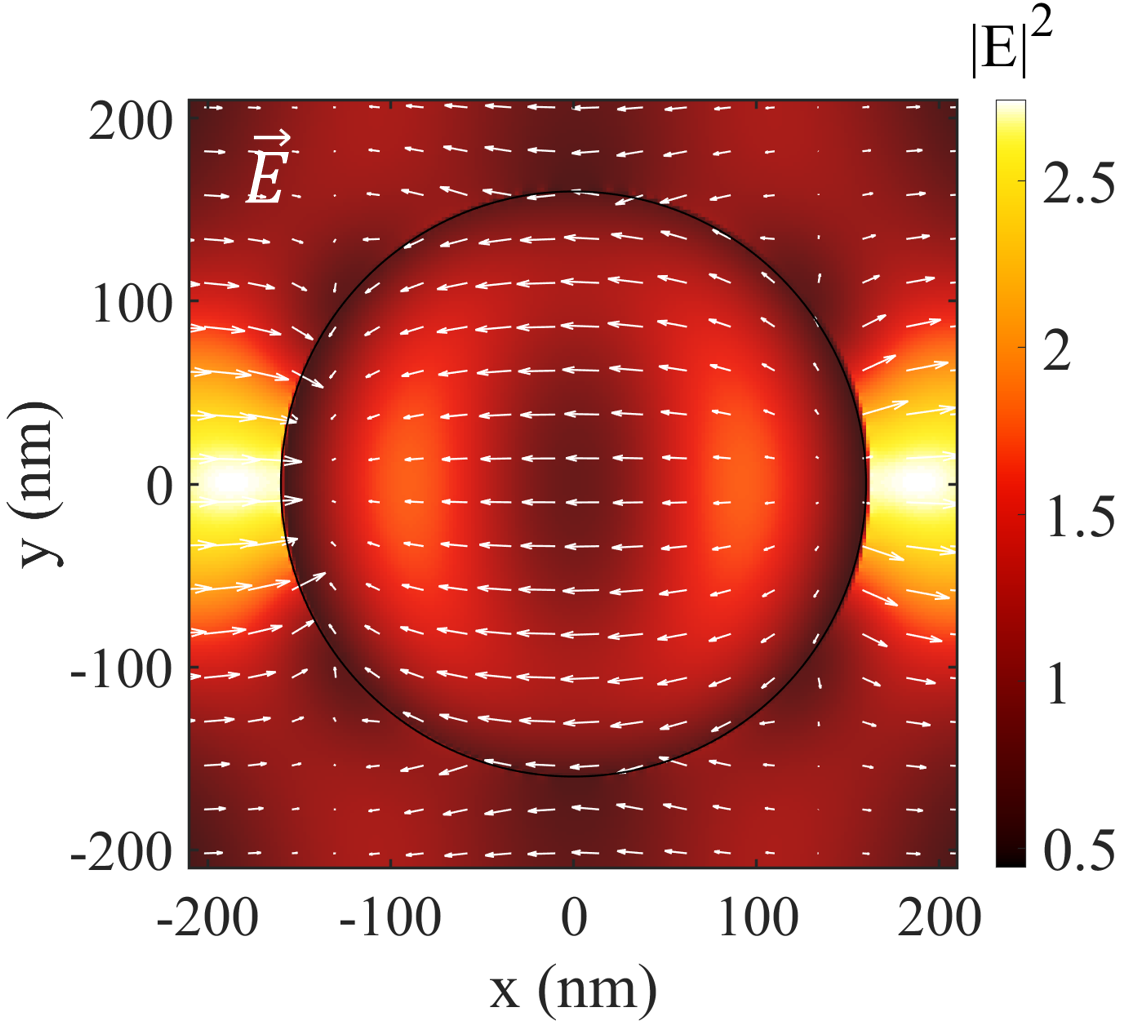}
(c)\includegraphics[width=0.306\linewidth]{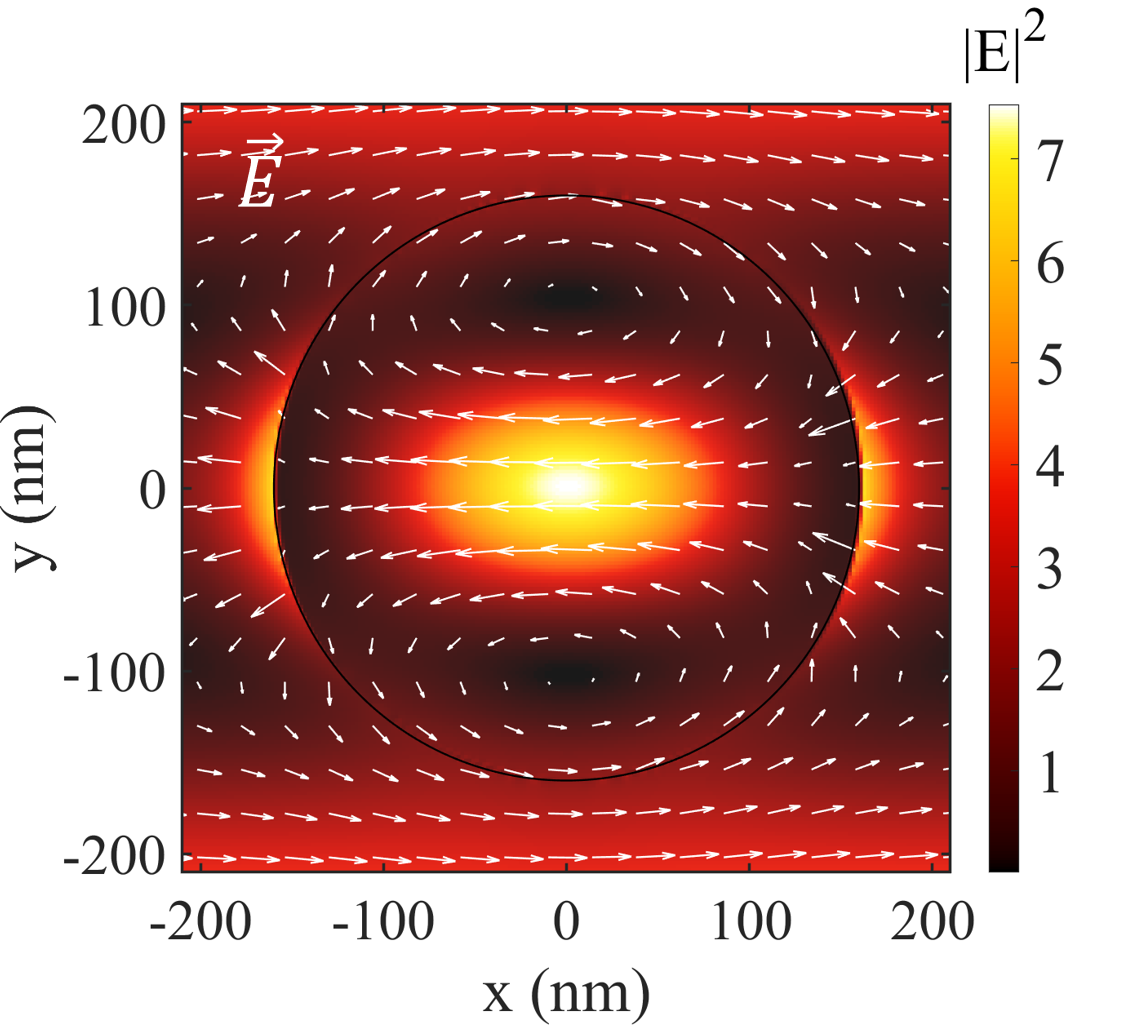}

\vspace{2mm}

(d)\includegraphics[width=0.305\linewidth]{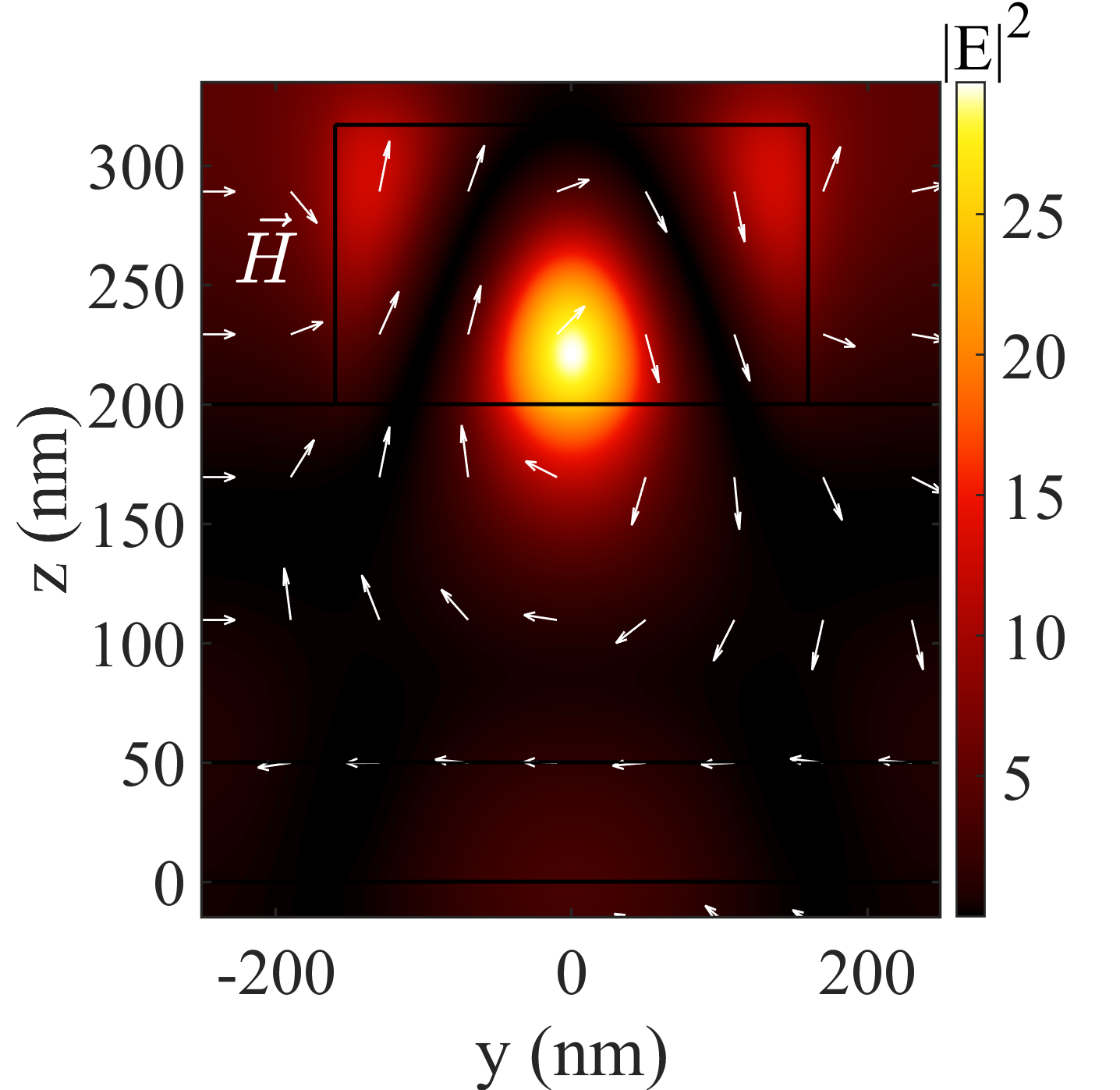}
(e)\includegraphics[width=0.305\linewidth]{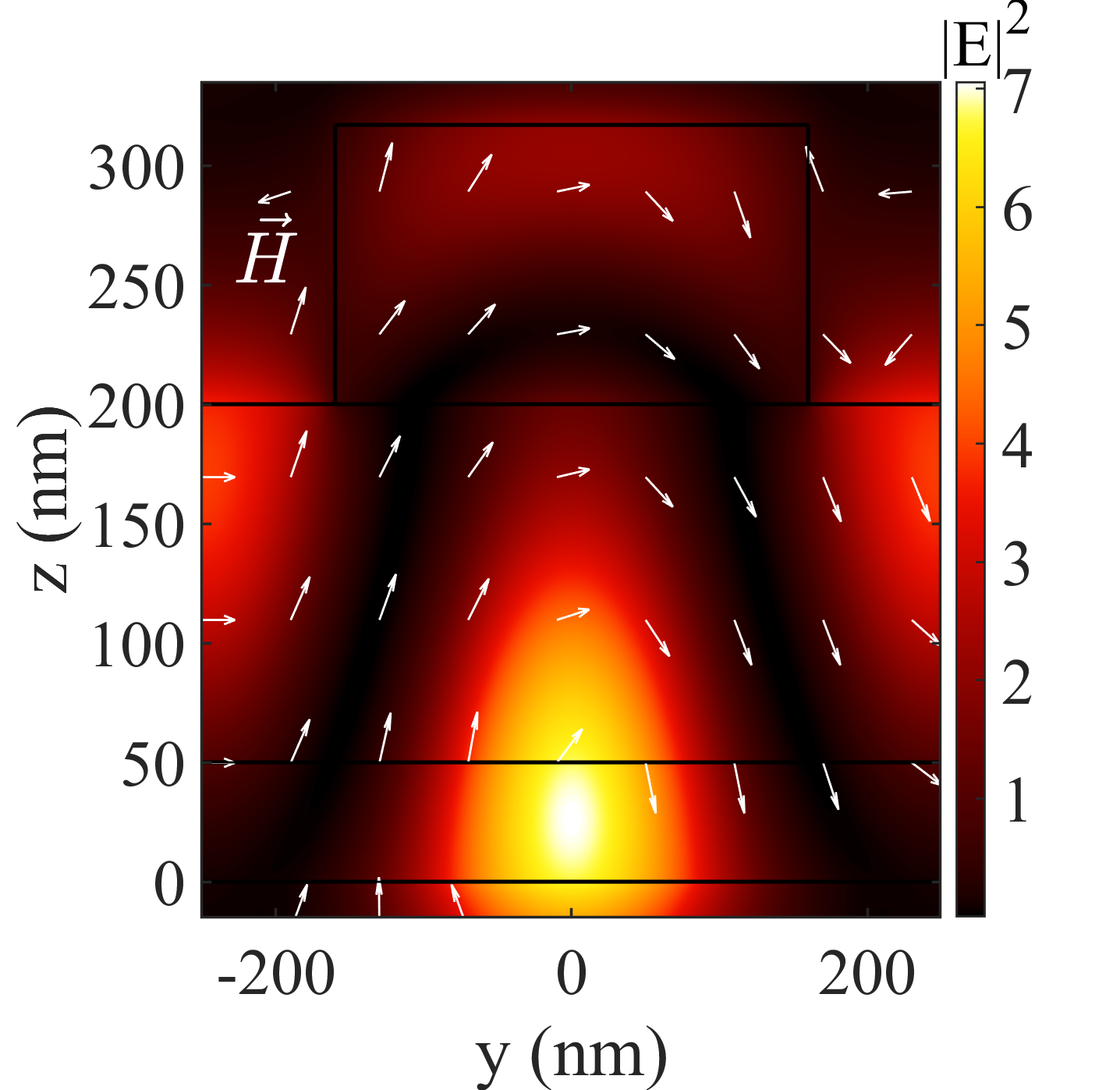}
(f)\includegraphics[width=0.305\linewidth]{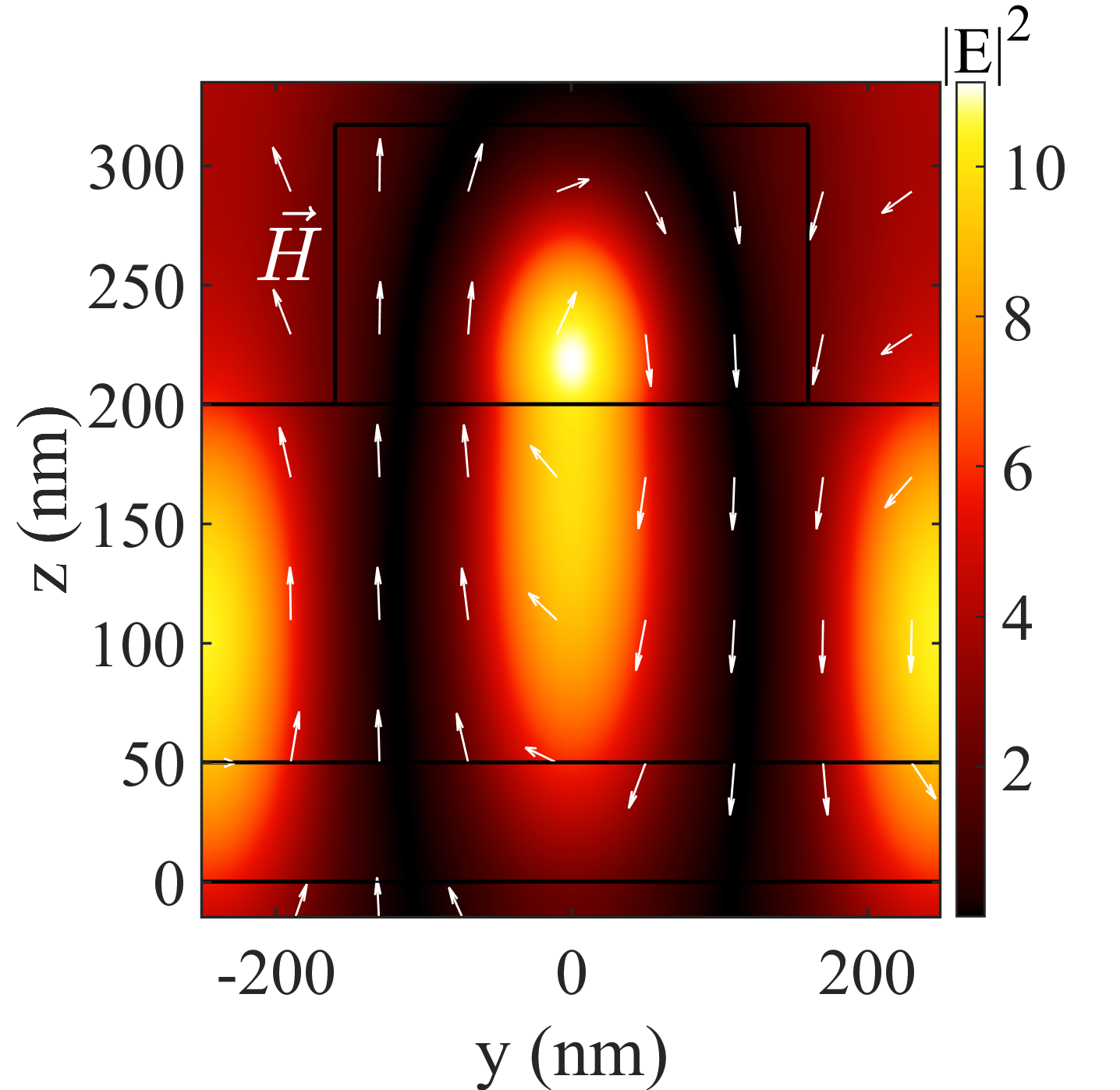}

\caption{
Electric-field distributions for different spectral features.
Top row (a--c): $|E|^2$ in the $x$--$y$ plane at half the nanodisk height for
(a) the anapole state at $\lambda=825~\mathrm{nm}$,
(b) the magnetic dipole (MD) mode at $\lambda=941~\mathrm{nm}$, and
(c) the electric dipole (ED) mode at $\lambda=997~\mathrm{nm}$.
Bottom row (d--f): the corresponding $|E|^2$ distributions in the $yz$ cross-sectional plane through the nanodisk center for the same states:
(d) anapole state, (e) MD, and (f) ED.
In (a--c), the black circle outlines the nanoparticle (disk) and white arrows indicate the in-plane electric-field $\vec{E}$.
In (d--f), white arrows indicate the direction of the magnetic-field vector $\vec{H}$, and black lines outline the boundaries of the layered structure.
}
\label{fig:fields_xy_yz}
\end{figure*}

The characteristic spectral features corresponding to the aforementioned states are observed in the transmittance spectra, see Fig.~\ref{Fig: scheme and T}b,c. The most prominent and spectrally broad features are associated with localized modes of the nanodisks, whereas they are accompanied by narrower angle-dependent resonances originating from guided modes of the structure. The detailed analysis of the latter features is presented in Appendix~D and is not discussed here.

Magnetic (MD) and electric (ED) dipole modes manifest themselves as pronounced transmission minima in the experimental spectra, at wavelengths of about $956~\mathrm{nm}$ and $997~\mathrm{nm}$, respectively. In contrast, the anapole state appears as a local transmission maximum at $\lambda = 839~\mathrm{nm}$~(Fig.~\ref{Fig: scheme and T} b). The transmittance at this wavelength reaches approximately $89\%$, rather than $100\%$, due to Fresnel reflection at the air--garnet and substrate interfaces. This regime is accompanied by suppressed far-field radiation and strong energy localization inside the nanoparticle. The numerically calculated electromagnetic field distributions of these states, excited under normal incidence of linearly polarized light, clarify the physical origin of the observed resonances. In the simulations, the field profiles were analyzed at the corresponding wavelengths of $\lambda = 825~\mathrm{nm}$ for the anapole state, $\lambda = 941~\mathrm{nm}$ for the MD mode, and $\lambda = 997~\mathrm{nm}$ for the ED mode. Figures~\ref{fig:fields_xy_yz}(a--c) show the top-view distributions of the electric-field intensity $|E|^{2}$ and the in-plane electric-field components $(E_x,E_y)$ for the anapole, magnetic dipole (MD), and electric dipole (ED) states. Figures~\ref{fig:fields_xy_yz}(d--f) present the corresponding side-view distributions, where the arrows indicate the magnetic-field vector $\mathbf{H}$.

The anapole state is characterized by strong confinement of the electric field inside the nanodisk, accompanied by suppressed far-field radiation. In the spectra, it appears as a local transmission maximum associated with strong energy localization inside the nanoparticle. The electric-field distribution at the corresponding wavelength (Fig.~\ref{fig:fields_xy_yz}a) shows that the field is concentrated predominantly inside the nanoparticle, while the field outside the disk is substantially weaker. The in-plane electric-field components $(E_x,E_y)$ exhibit a vortex-like near-field pattern, indicating circulating displacement currents within the nanodisk. In the $yz$ cross-sectional plane (Fig.~\ref{fig:fields_xy_yz}d), the field remains strongly localized inside the silicon nanodisk, while its amplitude in the magnetic film and in the surrounding region is suppressed. In combination with the transmission maximum and the multipolar analysis presented in Appendix~E, where the condition $\mathbf{P}+ik\mathbf{T}=0$ is satisfied, these characteristics unambiguously identify this spectral feature as the anapole state ~\cite{miroshnichenko2015nonradiating}.

For the magnetic dipole (MD) mode, the electric-field distribution differs qualitatively from the anapole state. In the $x$--$y$ plane (Fig.~\ref{fig:fields_xy_yz}b), the electric-field intensity $|E|^{2}$ inside the nanodisk is comparatively weak, while enhanced field regions are observed mainly near the disk boundaries. In the $yz$ cross-sectional plane (Fig.~\ref{fig:fields_xy_yz}e), the electric-field intensity exhibits a pronounced maximum inside the magnetic layer beneath the silicon nanodisk. This behavior indicates substantial penetration of the field into the garnet film, where the field amplitude remains high due to the large refractive index of the magnetic layer and the corresponding redistribution of the electromagnetic energy into this region. The in-plane electric-field vectors exhibit a spatially varying distribution without a pronounced preferential orientation. Since the MD resonance is primarily characterized by the magnetic-field response, the $|E|^{2}$ and $(E_x,E_y)$ maps presented here serve mainly as a qualitative illustration of the near-field structure associated with the MD excitation.

For the resonance attributed to the electric dipole (ED) mode, the electric field is strongly nonuniform and predominantly localized inside the nanodisk. The $|E|^{2}$ map in the $x$--$y$ plane (Fig.~\ref{fig:fields_xy_yz}c) exhibits a pronounced intensity maximum in the central region of the disk, elongated along the polarization direction of the incident light. The in-plane vectors are mainly aligned along this direction, while a noticeable electric field is also present outside the disk. In the $yz$ cross-sectional plane (Fig.~\ref{fig:fields_xy_yz}f), the ED resonance shows a pronounced electric-field maximum extending across the silicon nanodisk and into the underlying magnetic film, indicating substantial field penetration into the magnetic layer. This spatial pattern is consistent with a resonant response dominated by the electric dipole contribution.

Overall, the calculated near-field maps reveal clear qualitative differences between the anapole, MD, and ED spectral features. In the following, we focus on how these differences in the near-field distributions influence the enhancement of magneto-optical effects, in particular magneto-optical transmission modulation. The Faraday and Kerr effects are considered separately in Appendix~F, as the main focus of the present work is on the intensity-type magneto-optical response in transmission.

\subsection{Mode-enhanced magneto-optical intensity effect}

\begin{figure}[htb]
\centering
\includegraphics[width=0.99\linewidth]{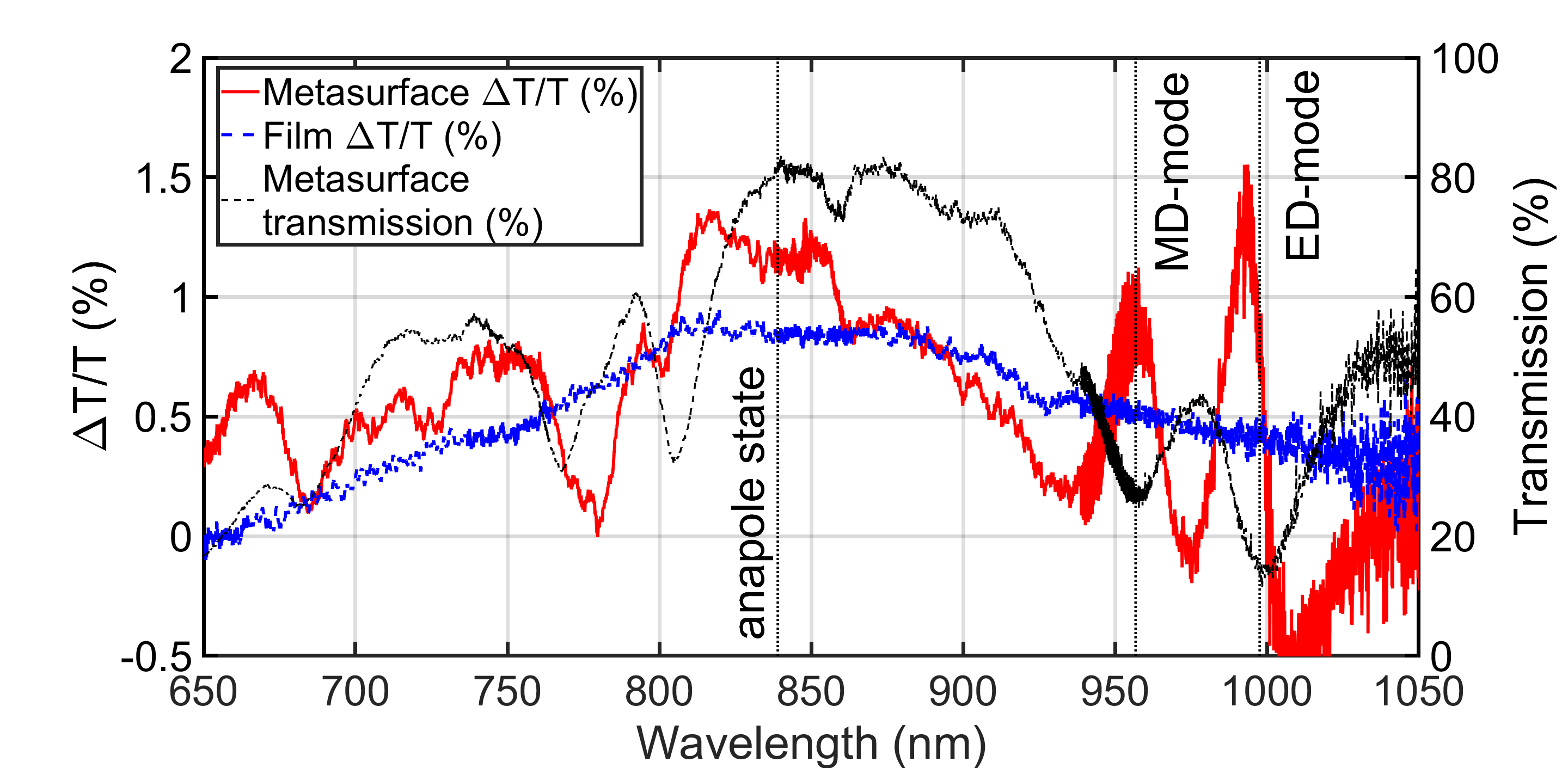}
\caption{Experimentally measured normalized magneto-optical intensity effect $\delta$ of the metasurface and the bare magnetic film under normal incidence of right-handed circularly polarized light.}
\label{fig:0_mcd}
\end{figure}
As the light localizes in the magnetic Dy:CeYIG film due to the optical response of the nanodisk array, magneto-optical transmission modulation arises under film magnetization. This intensity-type effect reveals itself as a magnetization-induced change in the absorption of circularly polarized light in a magnetized medium and also results in a difference of the transmittance and reflectance coefficients. We focus on the measurements of the transmittance variation due to this phenomenon, which in our experimental geometry can be detected by reversing the magnetization direction at fixed circular polarization:

\begin{equation}
\delta = \frac{\Delta T}{T} =
\frac{T(\sigma^{+},+M_z)-T(\sigma^{+},-M_z)}
{T(\sigma^{+},+M_z)+T(\sigma^{+},-M_z)},
\end{equation}
where $\sigma^{+}$ and $\sigma^{-}$ denote the right- and left-handed circular polarizations of light, respectively, so that $\delta(\sigma^{+})=-\delta(\sigma^{-})$.

\begin{figure*}[htb]
\centering
(a)\includegraphics[width=0.37\linewidth]{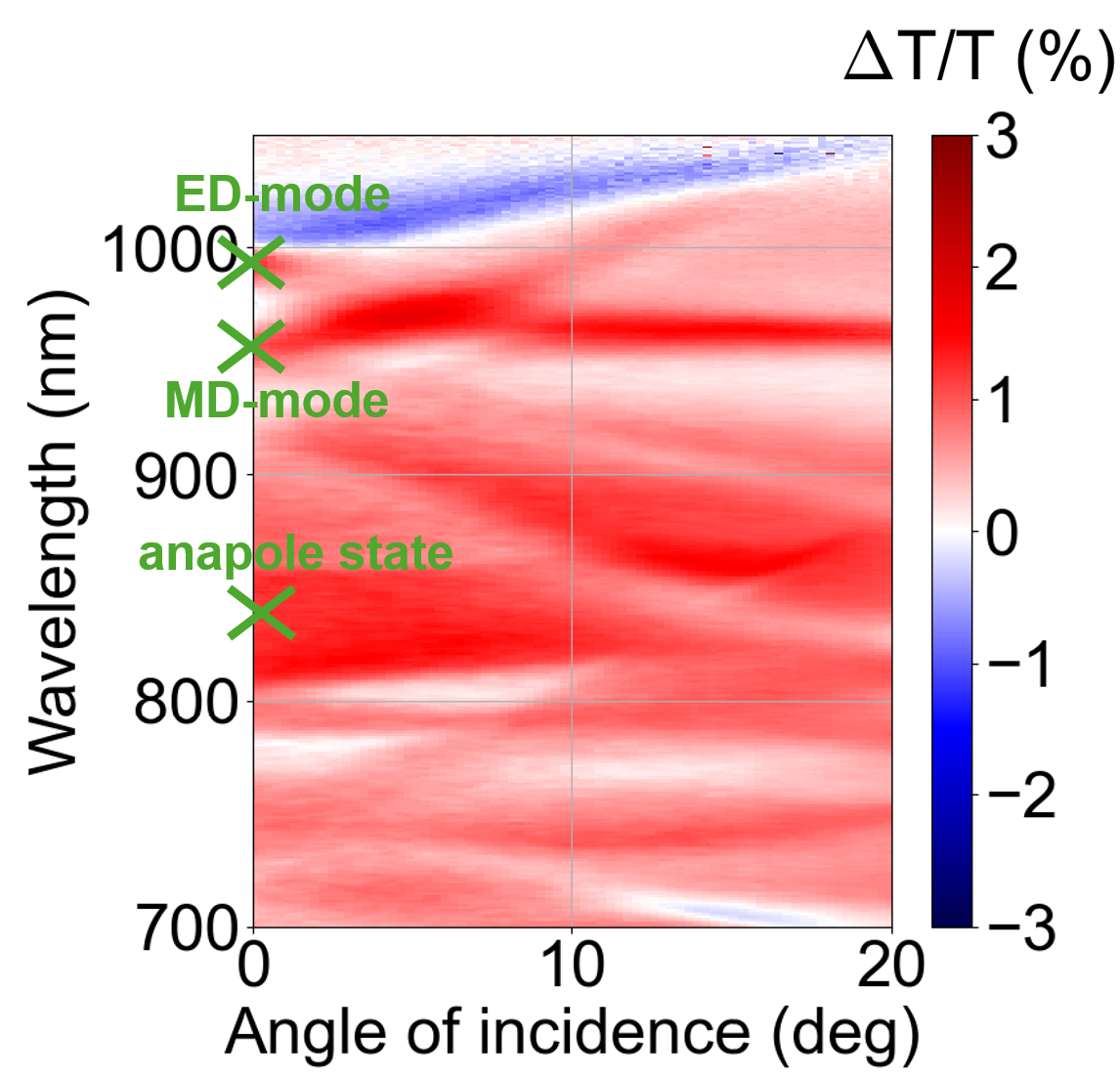}
(b)\includegraphics[width=0.37\linewidth]{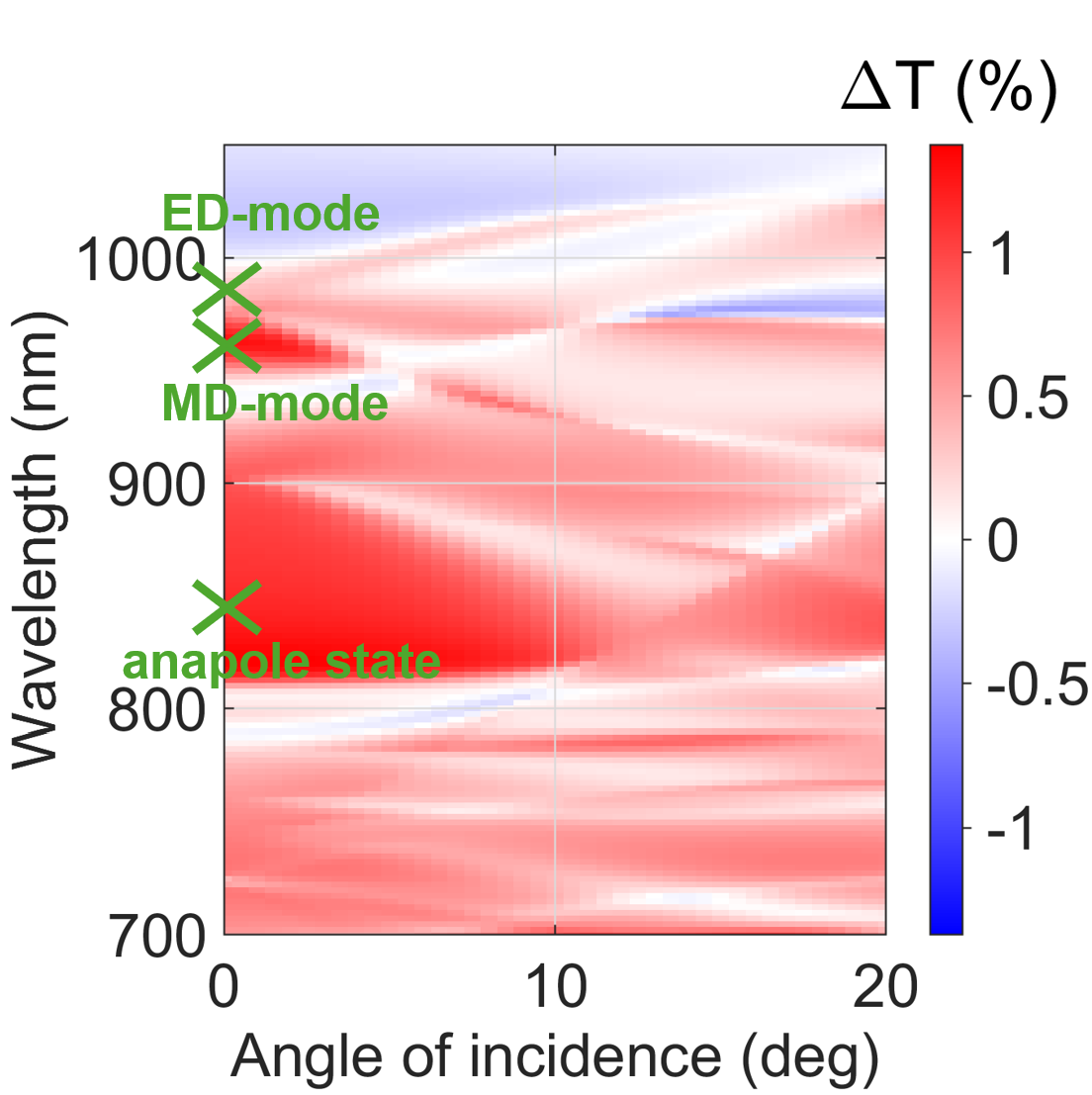}
\caption{Angle-resolved magneto-optical transmission modulation of the sample under circularly polarized excitation:
(a) experimental normalized signal $\delta$, (b) calculated unnormalized signal $\Delta T$}
\label{fig:mcd}
\end{figure*}

Figure \ref{fig:0_mcd}a shows the experimentally measured $\delta(\lambda)$ spectra for the metasurface and the same bare film Dy:CeYIG without nanodisk pattern. While the film itself has a smooth spectral $\delta(\lambda)$ dependency, the metasurface exhibits several resonant features, where the values $\delta$ are significantly enhanced. 

The most notable enhancement is observed at the wavelengths corresponding to the electric and magnetic dipole resonances. In particular, in the vicinity of the magnetic dipole (MD) mode the normalized transmission modulation reaches values of about 1\%, while near the electric dipole (ED) mode it increases up to approximately 1.5\%. For comparison, the bare magnetic film in the same spectral range exhibits a normalized transmission modulation of about 0.5\%. This indicates that both ED and MD resonances provide a noticeable enhancement of the magneto-optical response compared to the bare film.

The normalized magneto-optical intensity effect of the metasurface at its maximum, observed in the spectral region around $\lambda = 829$~nm, exceeds that of the bare film by approximately 32\%. It is significant that the maximum value of the normalized transmission modulation is observed in the same spectral region where the transmission of the metasurface reaches its maximum. Thus, the strongest normalized magneto-optical response is observed in the spectral region previously identified as the anapole state.



The pronounced transmission-modulation response in the spectral region associated with the anapole state is consistent with the nature of this mode. As shown in Fig.~\ref{fig:fields_xy_yz}(a,d), the anapole state is characterized by strong electromagnetic-field localization inside the silicon nanodisk together with suppressed far-field radiation. Such a regime corresponds to efficient energy storage in the resonant mode and reduced radiative losses, making the optical response of the structure more sensitive to absorption-related effects. Since this intensity effect originates from magnetization-induced changes in absorption for circularly polarized light, this provides a natural explanation for the pronounced transmission-modulation feature observed in the anapole spectral region. The local transmission maximum observed in the same range is therefore regarded as a characteristic spectral signature of the anapole state rather than as the direct origin of the enhanced magneto-optical response.

Figure~\ref{fig:mcd} presents the angle-resolved magneto-optical intensity effect under circularly polarized excitation: the experimentally measured normalized signal $\delta$ (a) and the modelled magneto-optical response evaluated as $\Delta T = T(+M)-T(-M)$ (b). Here, the angle of incidence is measured from the sample normal, and in both the experiment and the modelling it is varied in the $xz$ plane.

Despite this difference in representation, the experimental and calculated maps exhibit the same qualitative behavior. In both cases, the strongest magneto-optical response is concentrated within the same spectral window, with the largest values observed predominantly in the $\sim 800$--$900$~nm range. This interval coincides with the high-transmission region identified from the transmission maps discussed above and associated with the anapole spectral region. The agreement between experiment and modelling indicates that the enhancement of the magneto-optical intensity effect is governed by the resonant excitation associated with the anapole spectral region.

The angle-resolved maps further show that the enhanced transmission-modulation response is not restricted to normal incidence but persists over a broad range of incidence angles within the anapole spectral region. This result is important because it demonstrates that the resonantly enhanced magneto-optical response is robust with respect to angular variation and is not limited to a narrow excitation geometry. Outside this spectral window, the signal becomes weaker and less structured, which further supports its resonant origin. The broad angular range over which the enhancement is maintained constitutes an important advantage of the structure, indicating that the magneto-optical amplification does not require fine tuning of the incidence angle and remains pronounced over an extended angular interval.

\section{Conclusion}

In this work, we have experimentally and numerically investigated an all-dielectric magneto-optical metasurface based on a bilayer magnetic garnet film (YIG/Dy:CeYIG) combined with a periodic array of silicon nanodisks. It is shown that the optical resonant states of the structure, including electric and magnetic dipole responses of the nanodisks as well as an anapole state, lead to a pronounced spectral modification of the transmission modulation compared to a bare magnetic film of the same thickness.

In the spectral regions corresponding to the magnetic and electric dipole resonances of the silicon nanodisks, the experimentally measured magneto-optical intensity effect of the metasurface also exceeds that of the bare magnetic film. In particular, the normalized transmission modulation reaches approximately 1\% at the magnetic dipole resonance and about 1.5\% at the electric dipole resonance, while the magneto-optical intensity effect the reference film in the same wavelength ranges does not exceed $\sim 0.5\%$. This corresponds to an enhancement by a factor of about 2–3 relative to the bare film.

The strong transmission-modulation response is observed in the spectral region corresponding to the anapole state, where the metasurface transmission reaches a local maximum. This regime is associated with suppression of radiation into the far field and with strong accumulation of electromagnetic energy inside the nanodisk, in agreement with the calculated near-field distributions. In the experiment, the maximum normalized transmission modulation of the metasurface exceeds that of the reference film by more than $30\%$. This enhanced response is preserved over a broad range of incidence angles, indicating that the magneto-optical amplification in the spectral region associated with the anapole state is weakly dependent on the excitation angle.

A comparison between experimental measurements, RCWA simulations, and analytical mode approach confirms that the enhancement of the magneto-optical intensity effect correlates with the excitation of the MD and ED modes of the structure, with the spectral feature associated with the anapole state, and with the degree of electromagnetic field localization in the magnetic layer. The obtained results demonstrate that anapole regimes in all-dielectric magneto-optical metasurfaces provide a promising route to achieve enhanced magneto-optical intensity responses while maintaining high optical transmission.

This research was funded by the Russian Science
Foundation, project N 24-42-02008.

\section{Appendix}

\subsection{Fabrication}

The magnetic garnet films were fabricated by pulsed laser deposition on a fused-silica ($\mathrm{SiO_2}$) substrate. A 50-nm-thick yttrium iron garnet (YIG) layer was first deposited and served as a seed layer to promote the crystallization of the upper magnetic film. Subsequently, a 150-nm-thick cerium-substituted dysprosium iron garnet (Dy:CeYIG) layer with aluminum doping was grown on top. The YIG layer was deposited at a substrate temperature of 400~$^\circ$C and an oxygen pressure of 10~mTorr, followed by rapid thermal annealing at 900~$^\circ$C in an oxygen atmosphere. The Dy:CeYIG film was deposited at a substrate temperature of 750~$^\circ$C and an oxygen pressure of 5~mTorr using alternating ablation from garnet targets with different compositions. After the deposition of the magneto-optical layers, an amorphous silicon film with a thickness of 117~nm was deposited by plasma-enhanced chemical vapor deposition. A two-dimensional array of silicon nanodisks was then fabricated using electron-beam lithography with a negative-tone HSQ resist, followed by reactive ion etching.

\subsection{Experimental measurements}
\begin{figure}[htb]
\centering
\includegraphics[width=0.99\linewidth]{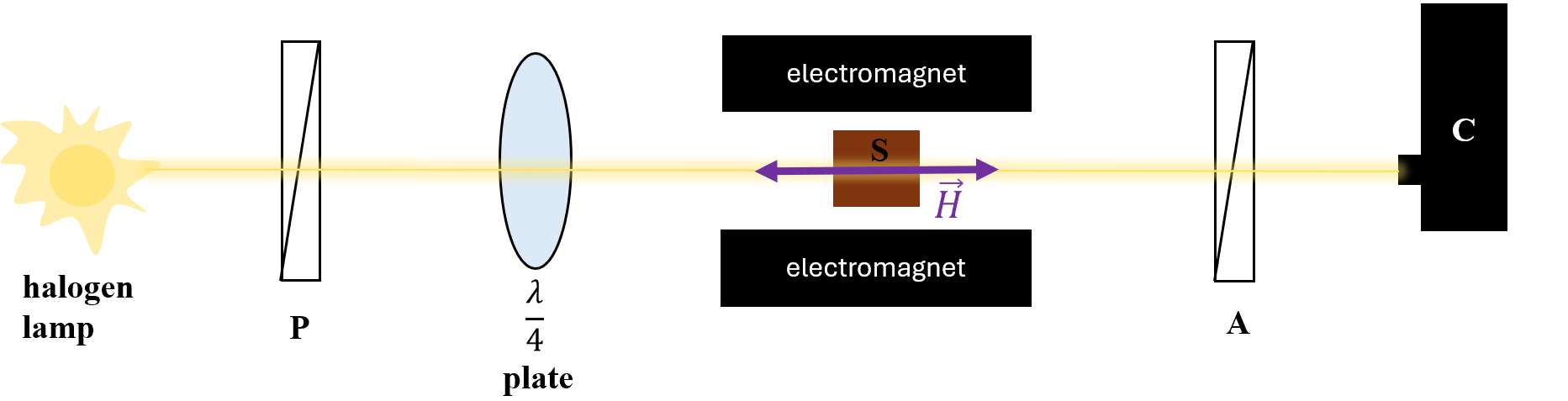}
\caption{Experimental setup}
\label{fig:exp_setup}
\end{figure}

The experimental setup was designed to measure the magneto-optical response of the metasurface in transmission geometry under circularly polarized illumination (Fig.~\ref{fig:exp_setup}, where P denotes the polarizer, A the analyzer, $\lambda/4$ the quarter-wave plate, and C the spectrometer; additional lenses and apertures used for beam shaping are not shown in the schematic). A broadband halogen lamp was used as a light source. The incident beam was formed by an optical system including lenses and apertures, then passed through a linear polarizer followed by a quarter-wave ($\lambda/4$) plate to generate circularly polarized light. The radiation was focused onto the sample, and the transmitted light was collected and directed to a CCD-based spectrometer in the spectral range of 200--1100~nm. Experimental parameters and data acquisition were controlled by a computer, and the recorded spectra were processed using dedicated software. 

The sample was positioned between the poles of an electromagnet providing a static magnetic field oriented normal to the sample plane. In this configuration, the magnetization vector was aligned parallel to the direction of light propagation, corresponding to the Faraday geometry. The magnetic field strength was set to 140 mT, which exceeds the saturation field of the sample (120 mT).

The transmitted spectra were recorded for opposite orientations of the magnetic field using a CCD-based spectrometer in the wavelength range of interest. The magneto-optical signal was evaluated as the difference between the transmission spectra obtained for opposite magnetization directions. This measurement scheme enables reliable detection of magnetization-induced transmission modulation.

\subsection{Optical and magneto-optical properties of materials}

\begin{figure}[t]
    \centering

    \begin{minipage}{\linewidth}
        \centering
        \textbf{(a)}\\[0.00001cm]
        \includegraphics[width=0.75\linewidth]{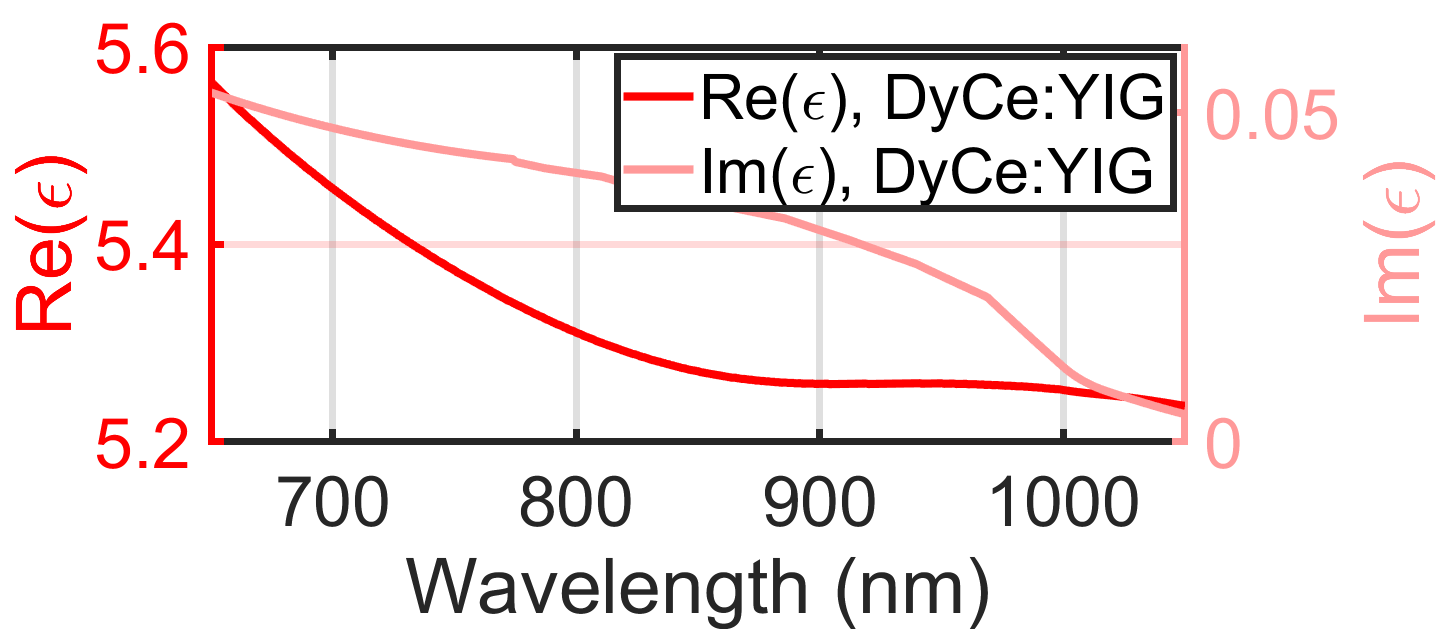}
    \end{minipage}
     \vspace{0.5cm}

    \begin{minipage}{\linewidth}
        \centering
        \textbf{(b)}\\[0.00001cm]
        \includegraphics[width=0.75\linewidth]{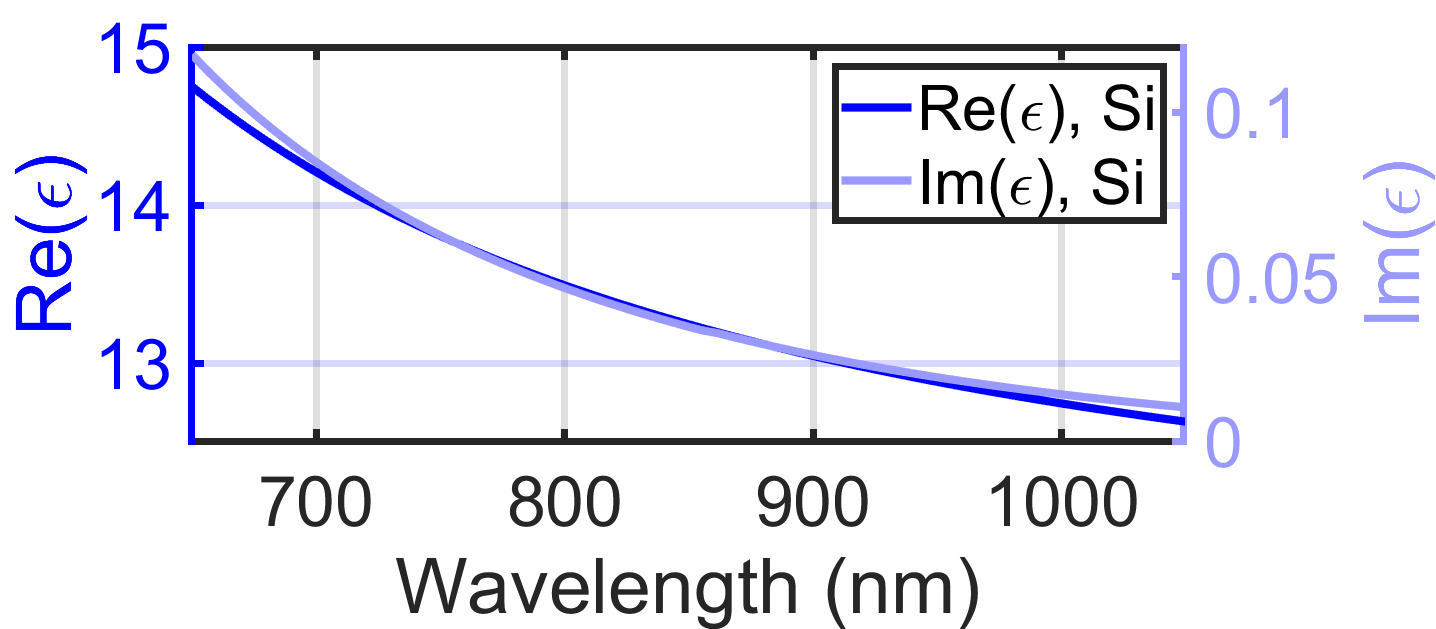}
    \end{minipage}

    \vspace{0.5cm}

    \begin{minipage}{\linewidth}
        \centering
        \textbf{(c)}\\[0.00001cm]
        \includegraphics[width=0.75\linewidth]{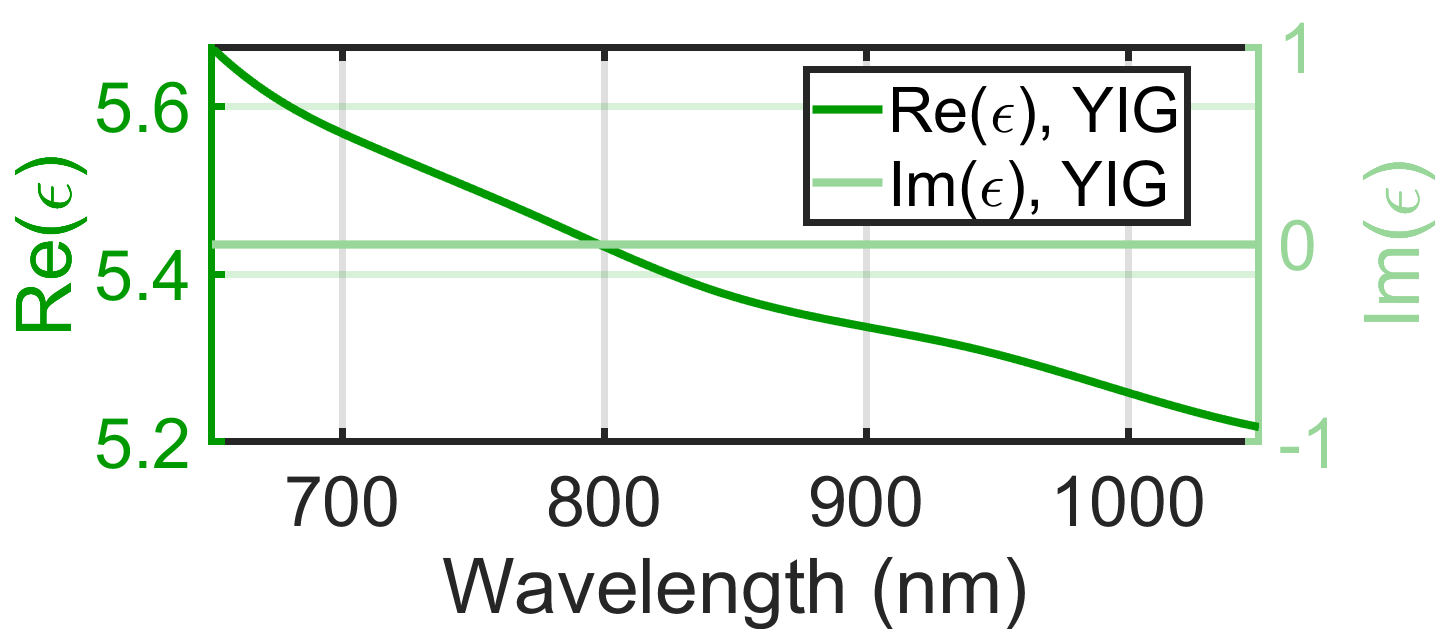}
    \end{minipage}
    \caption{Spectral dispersion of the complex dielectric permittivity for different materials:
    (a) Dy:CeYIG film, (b) silicon, and (c) YIG film}
    \label{fig:eps_dispersion}
\end{figure}

The spectral dispersion of the complex dielectric permittivity $\varepsilon(\lambda)$ for Dy:CeYIG, silicon, and YIG is shown in Fig.~\ref{fig:eps_dispersion}. The dielectric functions of the garnet materials were obtained experimentally, whereas the data for silicon were taken from~\cite{franta2017temperature}. These data were used as input parameters for the optical simulations.

\begin{figure}[t]
    \centering
    \includegraphics[width=0.9\linewidth]{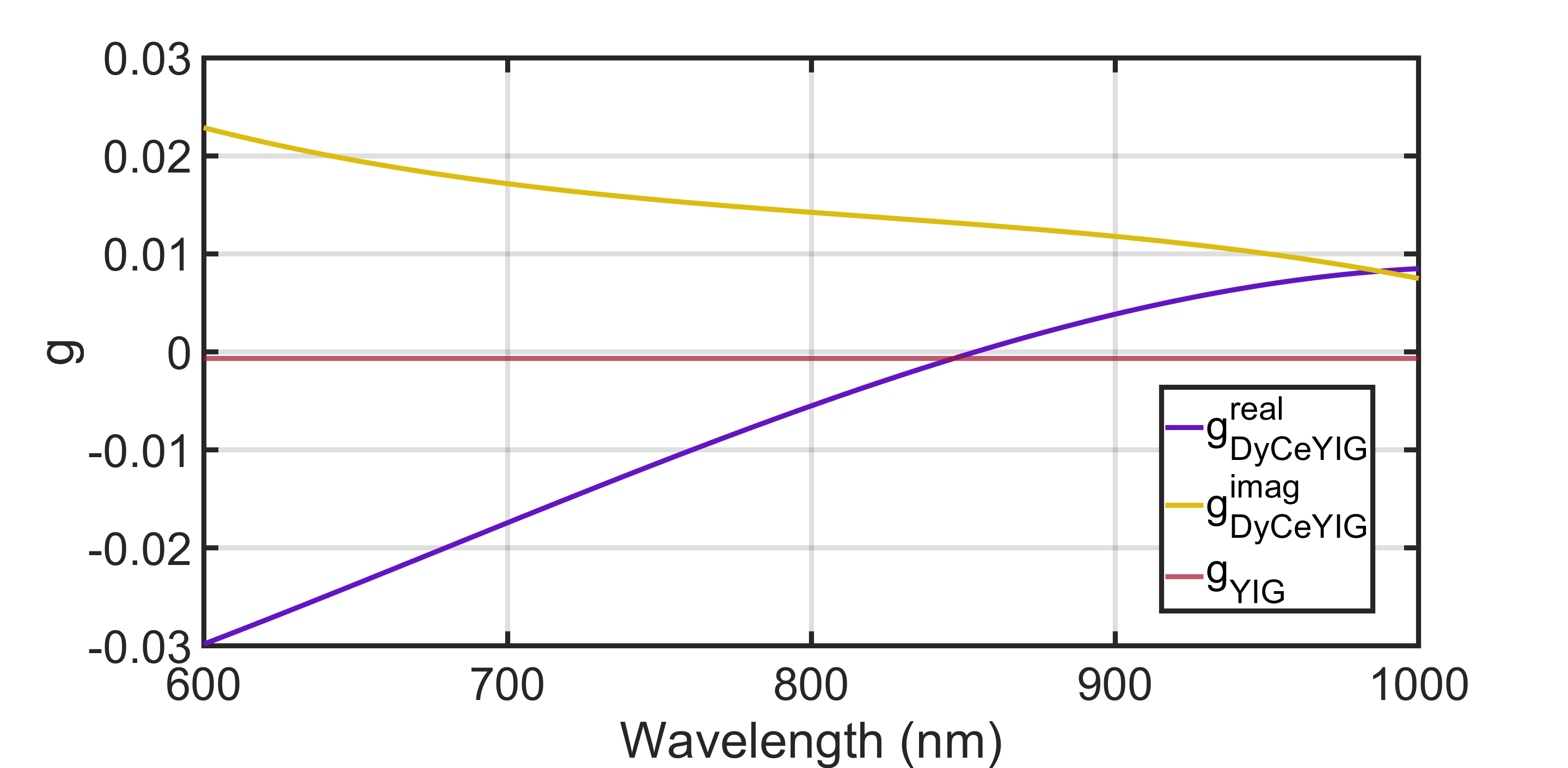}
    \caption{Spectral dispersion of the gyration $g(\lambda)$ used in the simulations}
    \label{fig:gyration}
\end{figure}

The wavelength dependence of the gyration $g(\lambda)$ is shown in Fig.~\ref{fig:gyration}. The gyration spectra of the garnet materials were also obtained experimentally and used in the magneto-optical calculations.

\subsection{Guided modes of the metasurface}

Experimentally measured angle-resolved transmission maps obtained under circularly polarized excitation reveal several pronounced resonance features (Fig.~\ref{Fig: scheme and T}b). To interpret these features, we use a guided-mode model in which the incident radiation couples to waveguide modes of the film via diffraction on the two-dimensional periodic structure. The incidence angle $\theta$ is varied in the $xz$ plane, so that the incident wavevector lies in this plane and its in-plane component is directed along the $x$ axis. Here, $\lambda$ denotes the free-space wavelength, and $\theta$ is the angle of incidence measured from the sample normal. In this framework, the resonance loci in the $(\lambda,\theta)$ plane are governed by the phase-matching condition between the in-plane wavevector component of the incident light and the modal propagation constant $\beta$, up to a reciprocal-lattice-vector contribution. For a rectangular lattice with periods $d_x$ and $d_y$, the condition is written as
\begin{equation}
\beta^{2}=\left(k_{0}\sin\theta+\frac{2\pi m}{d_{x}}\right)^{2}+\left(\frac{2\pi n}{d_{y}}\right)^{2}, 
\qquad k_{0}=\frac{2\pi}{\lambda},
\label{eq:phasematch}
\end{equation}
where $m$ and $n$ are integers defining the diffraction order along the $x$ and $y$ directions.

The modal dispersion $\beta(\lambda)$ is evaluated within an effective asymmetric slab-waveguide model, 
where a core layer of thickness $d$ and permittivity $\varepsilon_2$ is bounded by semi-infinite claddings 
with permittivities $\varepsilon_1$ and $\varepsilon_3$. 
Neglecting gyrotropy, the dispersion relations for the TE and TM modes take the standard form~\cite{yariv1984yeh}:

\begin{equation}
\begin{aligned}
-\,p_{2}d
&+\tan^{-1}\!\left[
\left(\frac{\varepsilon_{2}}{\varepsilon_{1}}\right)^{r}
\frac{p_{1}}{p_{2}}
\right] \\
&+\tan^{-1}\!\left[
\left(\frac{\varepsilon_{2}}{\varepsilon_{3}}\right)^{r}
\frac{p_{3}}{p_{2}}
\right]
=-N\pi, \\
&r=\{0~(\mathrm{TE}),\,1~(\mathrm{TM})\}.
\end{aligned}
\label{eq:slab_TE_TM}
\end{equation}
with
\begin{equation}
\begin{aligned}
p_{1}&=\sqrt{\beta^{2}-\varepsilon_{1}k_{0}^{2}},\\
p_{2}&=\sqrt{\varepsilon_{2}k_{0}^{2}-\beta^{2}},\\
p_{3}&=\sqrt{\beta^{2}-\varepsilon_{3}k_{0}^{2}}.
\end{aligned}
\label{eq:p_def}
\end{equation}
Here, $N$ is an integer specifying the mode order along the $Oz$ direction. In the present model, the modal dispersion is evaluated within the slab-waveguide approximation neglecting gyration, so Eq.~\eqref{eq:slab_TE_TM} is used for both TE and TM modes. In the considered geometry, magneto-optical contributions to the waveguide-mode dispersion are quadratic in the gyration parameter and therefore remain very small for both polarizations. They are thus neglected in the present spectral analysis. For completeness, the full expression including the gyration-induced correction for one of the polarizations is given in~\cite{krichevsky2021silicon}.

\begin{figure}
    
\includegraphics[width=0.99\linewidth]{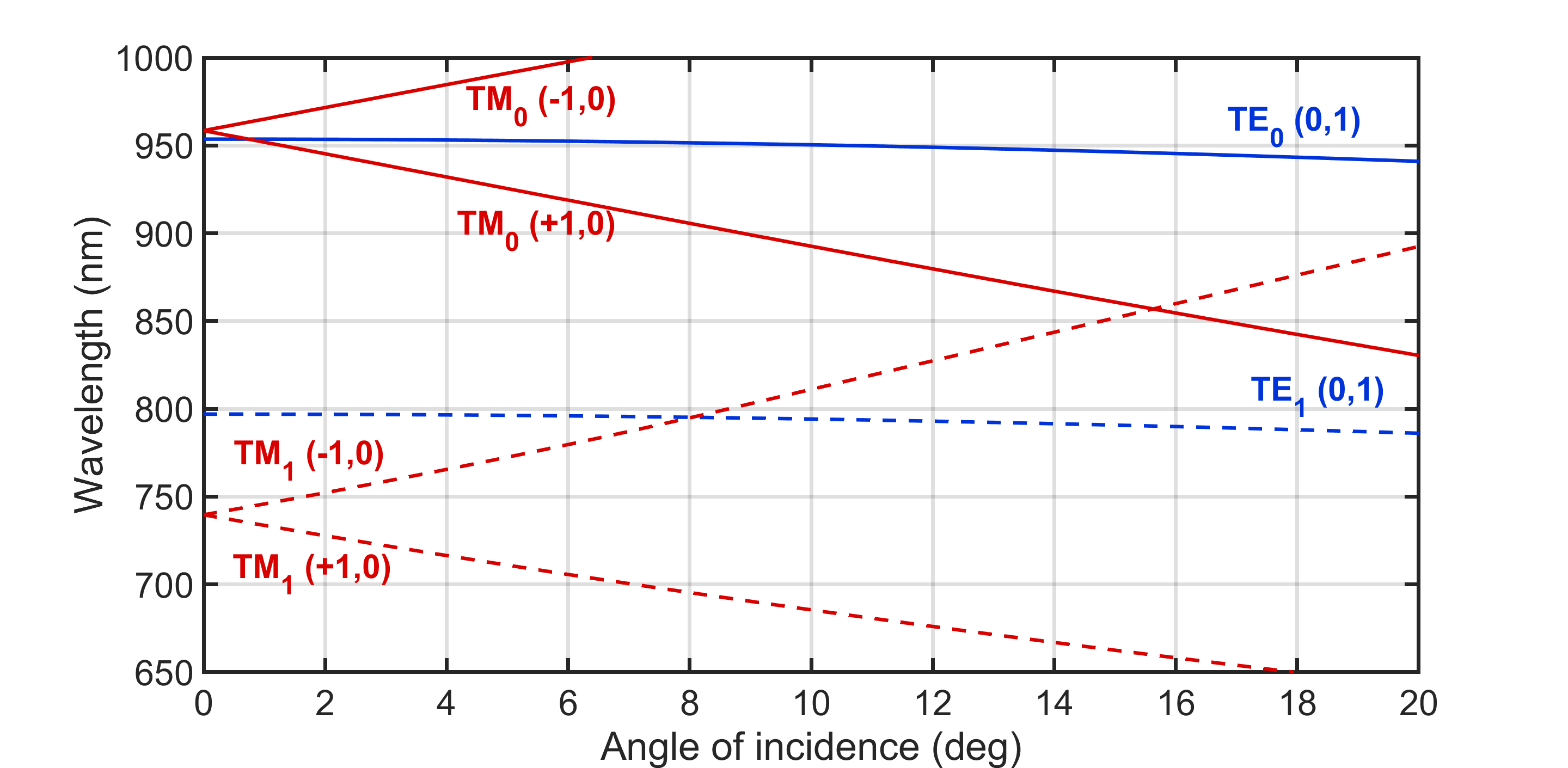}

\caption{ Calculated modal dispersion (solutions of the dispersion equation).}
\label{fig:angle_resolved_T_160}
\end{figure}

Solving Eqs.~\ref{eq:phasematch}--\ref{eq:p_def} yields the predicted resonance positions, which are compared with the experiment and with the RCWA simulations (Fig.~\ref{Fig: scheme and T}b,c), while the corresponding calculated modal dispersion is shown in Fig.~\ref{fig:angle_resolved_T_160}. For a concise quantitative comparison, the resonance wavelengths extracted for the four guided modes are summarized in Table~\ref{tab:guided_modes}. Overall, the RCWA simulations reproduce the set of experimentally observed resonances and support their interpretation in terms of TE/TM guided modes predicted by the analytical dispersion relations. The remaining differences between the analytical model and both experiment and RCWA simulations are mainly due to the fact that the analytical treatment is based on the empty-lattice approximation. In turn, the small discrepancies between the RCWA results and the experimental data can be attributed to structural imperfections of the fabricated sample.
\begin{table}[htb]
\caption{Resonance wavelengths of the guided modes obtained from the analytical model, experiment, and RCWA simulations.}
\label{tab:guided_modes}
\begin{ruledtabular}
\begin{tabular}{lccc}
Mode & Analytical model (nm) & Experiment (nm) & RCWA (nm) \\
\hline
TM$_0$ & 958 & 1000 & 1000 \\
TE$_0$ & 953 & 910  & 940  \\
TE$_1$ & 797 & 805  & 805  \\
TM$_1$ & 739 & 770  & 790  \\
\end{tabular}
\end{ruledtabular}
\end{table}

\subsection{Condition for excitation of the radiationless anapole state}

The electric and toroidal dipole moments were calculated using the standard expressions for the current Cartesian multipoles given in~\cite{miroshnichenko2015nonradiating}.

\begin{figure}[htb]
\centering
\includegraphics[width=0.99\linewidth]{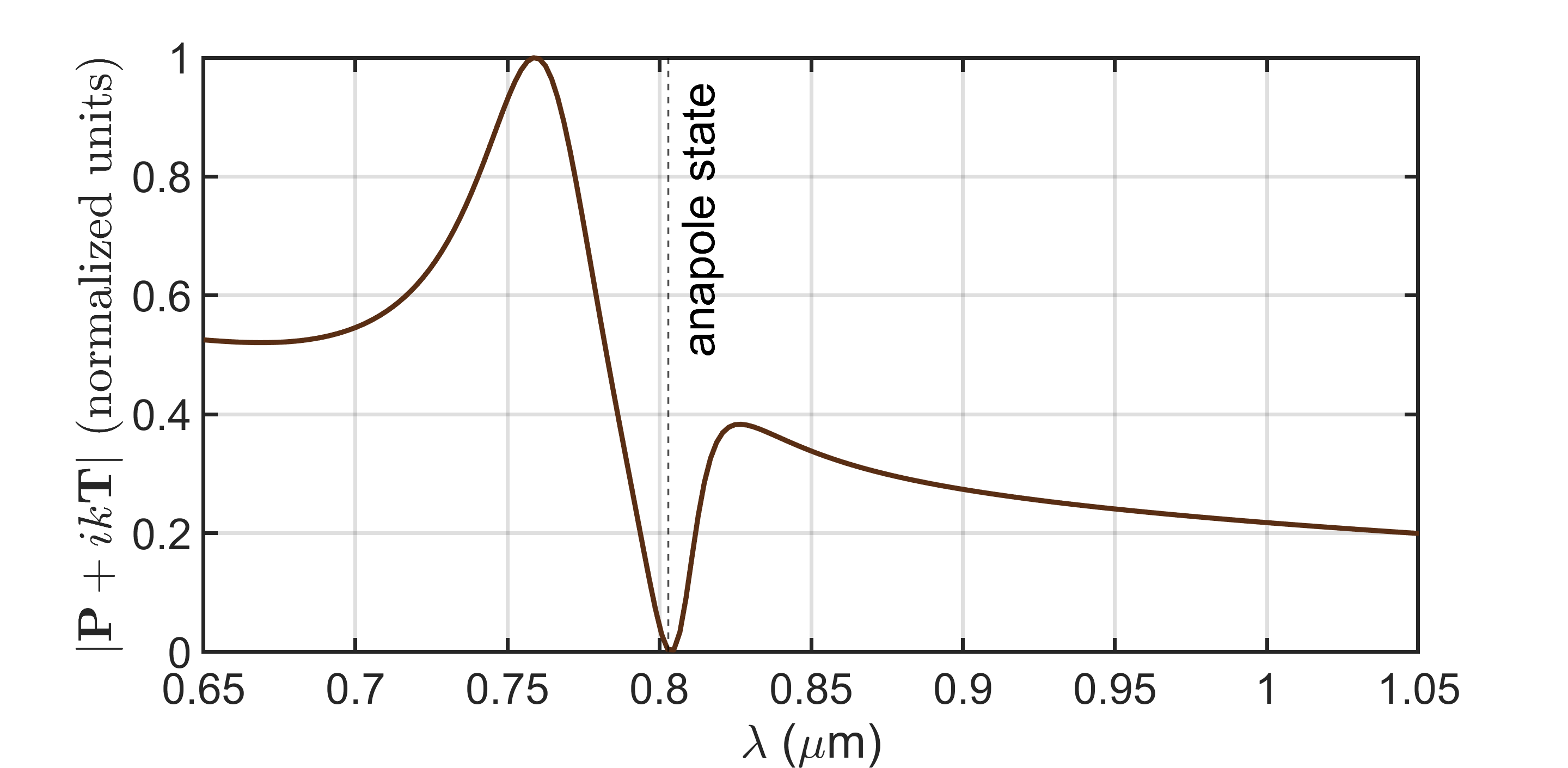}
\caption{Spectral dependence of the normalized modulus $|\mathbf{P} + ik\mathbf{T}|$, characterizing the compensation between the electric dipole moment and the toroidal dipole moment. The dashed vertical line indicates the anapole-state wavelength.}
\label{fig:anapolestate}
\end{figure}

Figure~\ref{fig:anapolestate} shows the spectral dependence of the normalized modulus $|\mathbf{P} + ik\mathbf{T}|$. At the wavelength of $0.803~\mu\mathrm{m}$, this quantity becomes zero, which corresponds to the exact fulfillment of the anapole condition.

This value is close to the wavelength of $0.825~\mu\mathrm{m}$, at which the anapole state was identified from the spectral response and the electromagnetic field distribution. The small difference between these values is expected, since the multipole analysis was performed for an isolated particle, whereas the spectral response was calculated for the full structure, taking into account the influence of the substrate, the garnet film, and the electromagnetic coupling between the elements of the array.

\subsection{Faraday and Polar Kerr effects}

\begin{figure}[htb]
\centering
\includegraphics[width=0.99\linewidth]{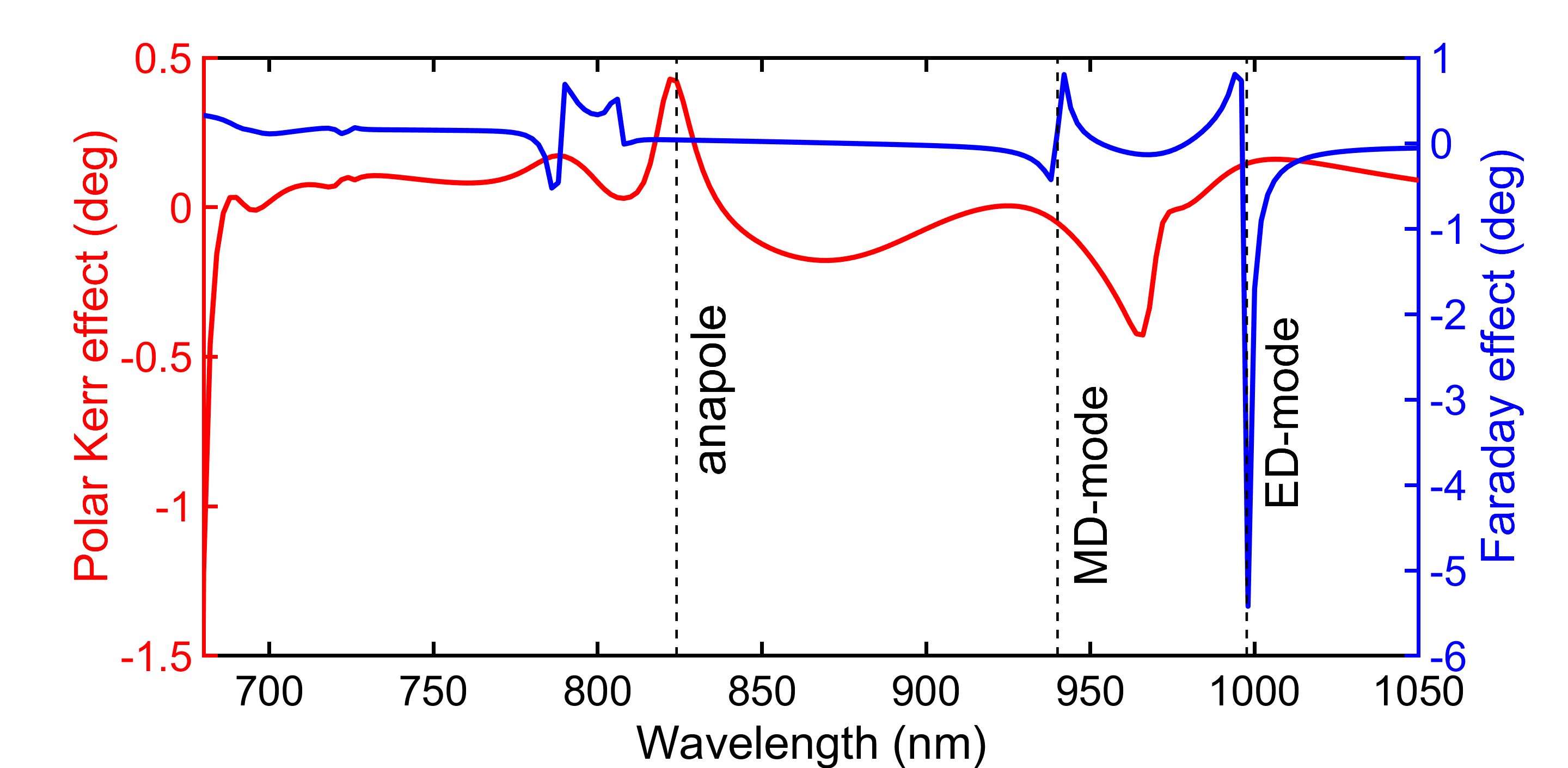}
\caption{Faraday rotation spectrum in transmission geometry and polar Kerr rotation spectrum in reflection geometry of the studied structure}
\label{Fig:FaradayKerr}
\end{figure}

For completeness, we briefly outline the conventional magneto–optical effects in our system. The Faraday effect, observed in transmission geometry, and the polar Kerr effect, observed in reflection geometry, represent direct manifestations of MO activity. However, in this work our focus is placed on magnetization-induced magneto-optical intensity effect rather than on these direct MO effects. The reason is that enhancement of Faraday or Kerr responses usually requires resonant conditions that increase the interaction time of light with the magnetic medium and thereby amplify the polarization rotation. In contrast, the anapole is not a resonance in this sense but an interference effect between electric and toroidal dipole moments; it does not extend the photon dwell time in the material but leads instead to strong field confinement inside the nanostructure, simultaneously with high overall transmission.

The calculated spectra illustrate this difference (Fig. ~\ref{Fig:FaradayKerr}). The Faraday rotation does not respond to the anapole state: it shows maxima at the magnetic dipole resonances, while no feature is seen at the anapole wavelength (where transmission is maximal and reflection minimal). In contrast, the Kerr rotation exhibits a feature at the anapole position; however, this is most likely related to the minimum in reflection rather than to a genuine enhancement of the MO response.

\newpage
\bibliography{apssamp}

\end{document}